\documentclass{article}
\usepackage{amsmath,epsfig}
\usepackage[preprint]{spconfa4}
\usepackage{xcolor}
\usepackage{cite}
\usepackage{subfigure}
\usepackage{multirow}
\usepackage{multicol}
\usepackage{caption}
\usepackage{textcomp}
\usepackage{xcolor}
\usepackage{hyperref}
\usepackage[misc]{ifsym}

\let\OLDthebibliography\thebibliography
\renewcommand\thebibliography[1]{
  \OLDthebibliography{#1}
  \setlength{\parskip}{0pt}
  \setlength{\itemsep}{0pt plus 0.3ex}
}

\begin{document}\sloppy

\def\x{{\mathbf x}}
\def\L{{\cal L}}

\title{Generative Compression for Face Video: A Hybrid Scheme}
%
\name{Anni Tang$^{\ast}$, Yan Huang$^{\ast}$, Jun Ling$^{\ast}$, 
Zhiyu Zhang$^{\ast}$, Yiwei Zhang$^{\ast}$, Rong Xie$^{\ast}$, 
Li Song\textsuperscript{$\ast$,$\dagger$ \Letter}
}
\address{
$^{\ast}$Institute of Image Communication and Network Engineering, Shanghai Jiao Tong University, China\\
$^{\dagger}$MOE Key Lab of Artificial Intelligence, AI Institute, Shanghai Jiao Tong University, China\\
\{memory97, 2250344200, lingjun, zhiyu-zhang, 6000myiwei, xierong, song\_li\}@sjtu.edu.cn
}

\maketitle

\begin{abstract}
As the latest video coding standard, versatile video coding (VVC) has shown its ability in retaining pixel quality. 
To excavate more compression potential for video conference scenarios under ultra-low bitrate, this paper proposes a bitrate-adjustable hybrid compression scheme for face video.
This hybrid scheme combines the pixel-level precise recovery capability of traditional coding with the generation capability of deep learning based on abridged information,
where Pixel-wise Bi-Prediction, Low-Bitrate-FOM and Lossless Keypoint Encoder collaborate to achieve PSNR up to 36.23 dB at a low bitrate of 1.47 KB/s.
Without introducing any additional bitrate, our method has a clear advantage over VVC under a completely fair comparative experiment, which proves the effectiveness of our proposed scheme. 
Moreover, our scheme can adapt to any existing encoder/configuration to deal with different encoding requirements, and the bitrate can be dynamically adjusted according to the network condition.
\end{abstract}
\section{Introduction}
\label{sec:intro}
In the context of COVID-19, more and more video calls are made, and face video compression technologies are becoming increasingly significant.
In the case of poor network quality, achieving stable and smooth video call with ultra-low bitrate is a hot research topic.
Existing methods for face video compression can be divided into two categories:
traditional coding methods and deep-learning-based methods.

Traditional video coding methods have the advantage of pixel-level precise recovery even for rapidly changing scenes, where VVC\cite{vvc} is the most advanced standard.
Compared with HEVC\cite{hevc}, VVC saves about 50\% of the bitrate while maintaining the similar visual quality.
However, these traditional methods represented by VVC do not discriminate video content and compress all videos in the same way although in fact, face video compression should have more potential.


Deep learning has the generation capability based on abridged information,
so it has great potential in face video compression.
Researchers have proposed some deep-learning-based methods\cite{Feng.2021.GCF,nvidia-maxine,facebook} to implement face video compression.
Oquab et al.\cite{facebook} designed a mobile-compatible architecture based on FOM\cite{fomm}, but bad results may appear with large attitude changes due to fixed reference frames.
Wang et al.\cite{nvidia-maxine} proposed an one-shot neural talking-head synthesis approach which utilizes one reference frame to generate a free-view output video, but the objective quality is not that satisfactory.
Feng et al.\cite{Feng.2021.GCF} proposed a video compression framework that utilizes 
pre-saved face frames as reference to reconstruct the target frame with fine details,
where additional bitrates are introduced
which leads to unfair bitrate comparison.
The above methods\cite{nvidia-maxine,facebook,Feng.2021.GCF,icassp} apply static reference frames,
so it's hard to guarantee good performance in case of scene switch or large attitude transformation, which leads to a failure in high fidelity recovery.
There also exists a generative compression method\cite{icassp} based on FOM\cite{fomm}, utilizing one raw frame as reference and adding the generated frames to the reference frame pool, which tends to cause error accumulation.
However, pixel-level accurate recovery is the strength of traditional coding methods.
Consequently, we come up with the idea of designing a hybrid scheme which combines traditional coding with deep generation.
Moreover, most of the existing AI-based video conference methods can hardly dynamically adjust the bitrate. 
For example, these methods\cite{nvidia-maxine,facebook,icassp,Feng.2021.GCF} can only adjust the bitrate through adjusting the number of keypoints despite the fact that different numbers of keypoints correspond to different models, further limiting their practicality.


\begin{figure*}[htb]
  \centering
  \includegraphics[width=0.95\linewidth]{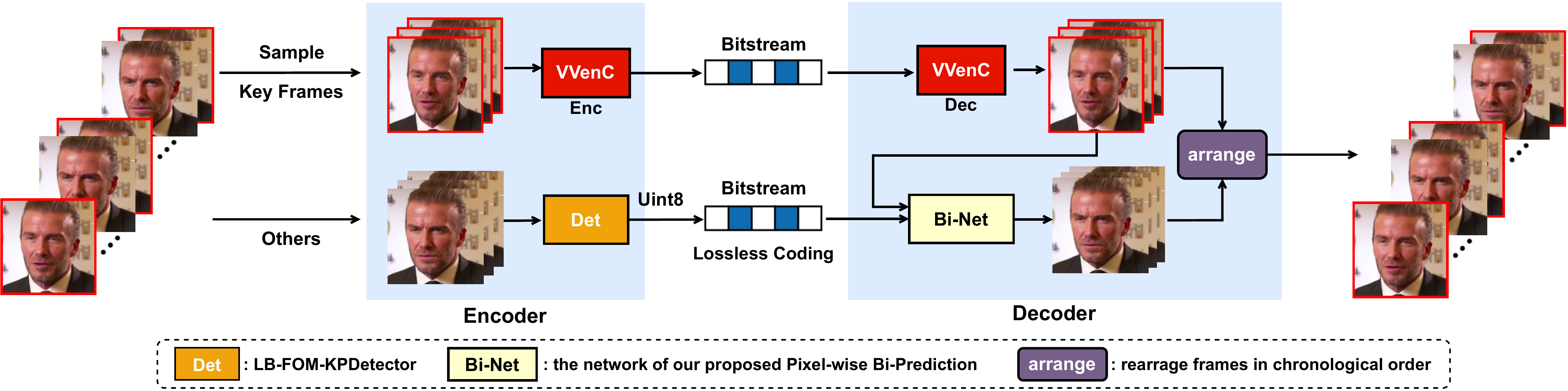}
  \caption{Overview of our method. 
  \textit{Sender}: (1) Sample key frames according to a certain sampling interval and use VVenC to encode them. (2) Extract keypoints of each non-key frame and encode them losslessly. (3) Transmit the VVenC and keypoint bitstream. \textit{Receiver}: (1) Decode the VVenC bitstream to get key frames. (2) Use the decoded key frames and the decoded keypoints of non-key frames to reconstruct the non-key frames via Bi-Net. (3) Rearrange all frames in chronological order.
  }
  \label{fig:pipeline}
\end{figure*}

\begin{figure}[tb]
  \subfigure[Forward Prediction.]{
      \includegraphics[width=0.45\linewidth]{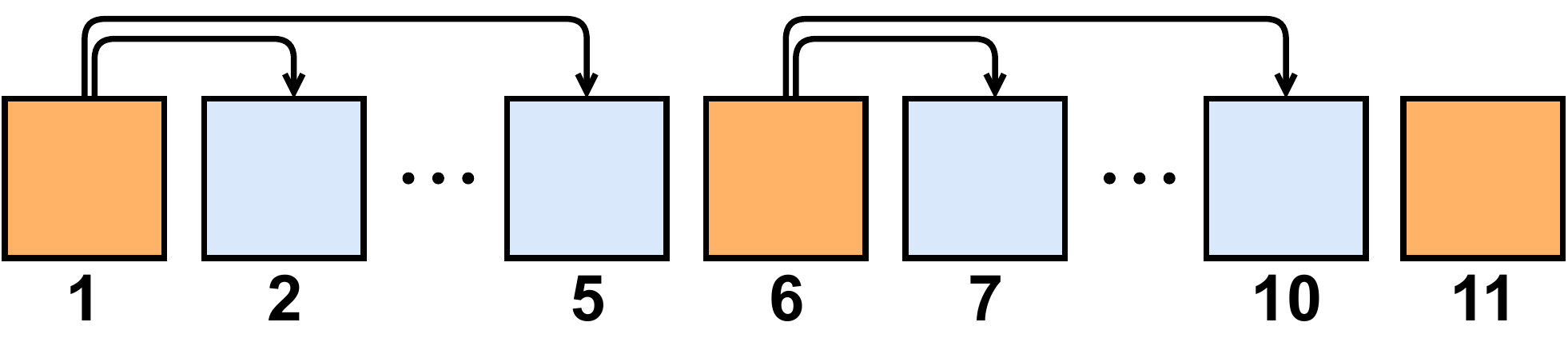}
  } 
  \hspace{0.1cm}
  \subfigure[Forward+Backward Prediction.]{
      \includegraphics[width=0.45\linewidth]{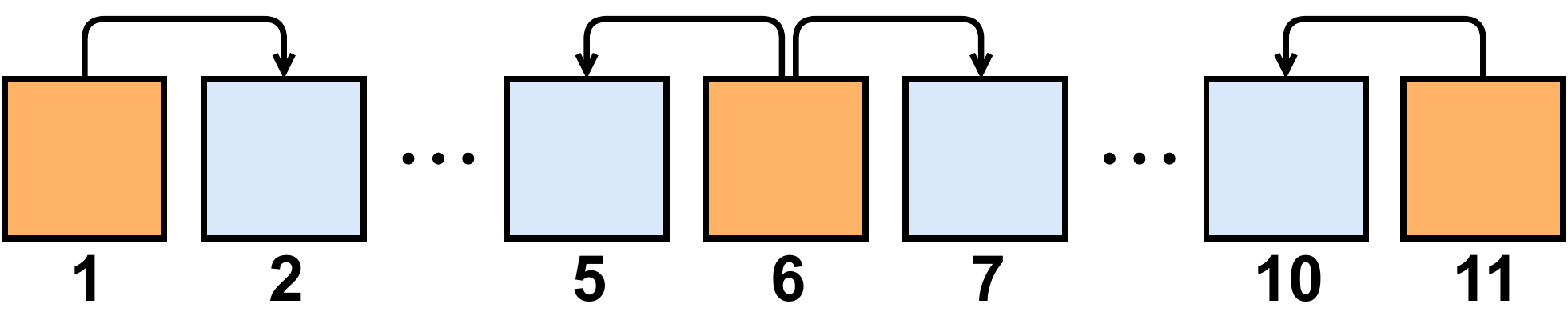}
  } \\
  \centering
  \subfigure[Pixel-wise Bi-Prediction.]{
      \label{fig:Pixel-wise_Bi-Prediction}
      \includegraphics[width=0.5\linewidth]{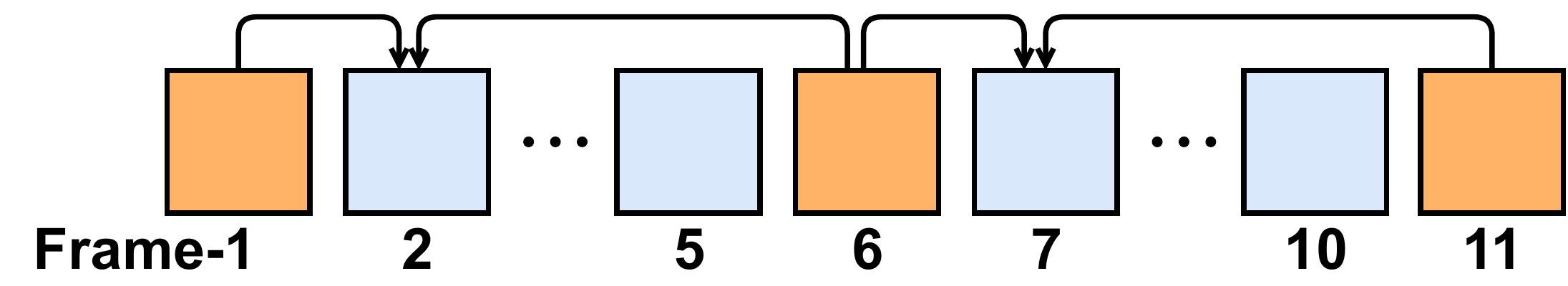}
  } \\
\caption{Three prediction methods. Assuming the sampling interval is 5, 
the orange blocks are key frames and the blue ones are non-key frames.
In (a), a non-key frame is predicted by the previous key frame.
In (b), a non-key frame is predicted by the key frame closest to it.
In (c), a non-key frame is predicted by two adjacent key frames jointly.
}
\label{fig:prediction_schemes}
\end{figure}
To tackle the aforementioned problems, we propose a bitrate-adjustable hybrid coding scheme.
It is a hybrid scheme because we sample the key frames according to a certain frequency, encode these key frames with traditional method such as VVC, and use deep generative method to reconstruct non-key frames according to the transmitted keypoints.
Combining the pixel-level recovery capability of traditional coding with 
the detail generation capability of deep learning, 
our proposed hybrid scheme is able to synthesize high-quality face
videos at ultra-low bitrate.
Meanwhile, the bitrate can be dynamically adjusted through two channels: the sampling interval of key frames or the compression ratio of key frames.
Thus, we can dynamically choose the right configuration according to the network condition, which is more practical than the aforementioned methods.

In our scheme, inspired by bidirectional prediction idea of traditional coding, we design a Pixel-wise Bi-Prediction method shown in Figure~\ref{fig:Pixel-wise_Bi-Prediction}, which uses two adjacent key frames and the keypoints of non-key frames to predict the non-key frames, greatly improving the smoothness of the generated video.
Specifically, we selected FOM\cite{fomm} as the basic generation model since it utilizes a sparse representation of human faces to implement face animation which costs a relatively-low bandwidth compared to \cite{wang2018videotovideo,wang2019fewshot,nirkin2019fsgan,bilayer}. 
Based on FOM\cite{fomm}, we propose a low bitrate version called LowBitrate-FOM (LB-FOM) to further lower the bitrate, taking the bitrate to one-twelfth of original FOM\cite{fomm} with little loss of quality, and design a lossless coding scheme to conduct further compression.
Overall, we assist traditional coding with deep generative methods to improve the detail retention and assist deep generative methods with traditional bidirectional prediction idea to improve smoothness, achieving the goal of generating smooth and detailed face video at ultra-low bitrate.

Our method does not introduce any additional bitrate, which means
there is not any prior information at the receiver such as pre-saved reference frames before transmission.
Thus, under a completely fair comparison, our method has an obvious advantage over VVC in face video compression, which proves the effectiveness of our proposed scheme. 
Moreover, our scheme applies to any existing encoder/configuration, to deal with different encoding requirements.
The contributions can be summarized as:
(1) We propose a hybrid compression scheme for face video which combines the advantages of traditional coding and deep learning, synthesizing high quality face video at ultra-low bitrate.
(2) A Pixel-wise Bi-Prediction method is proposed to generate smooth videos.
(3) We propose an LB-FOM which takes the bitrate to one-twelfth of original FOM\cite{fomm} with little loss of quality.
(4) Our scheme implements dynamically-adjustable bitrate and adapts to any existing encoder/configuration to deal with different encoding requirements.

\begin{figure*}[htb]
  \centering
  \includegraphics[width=0.95\linewidth]{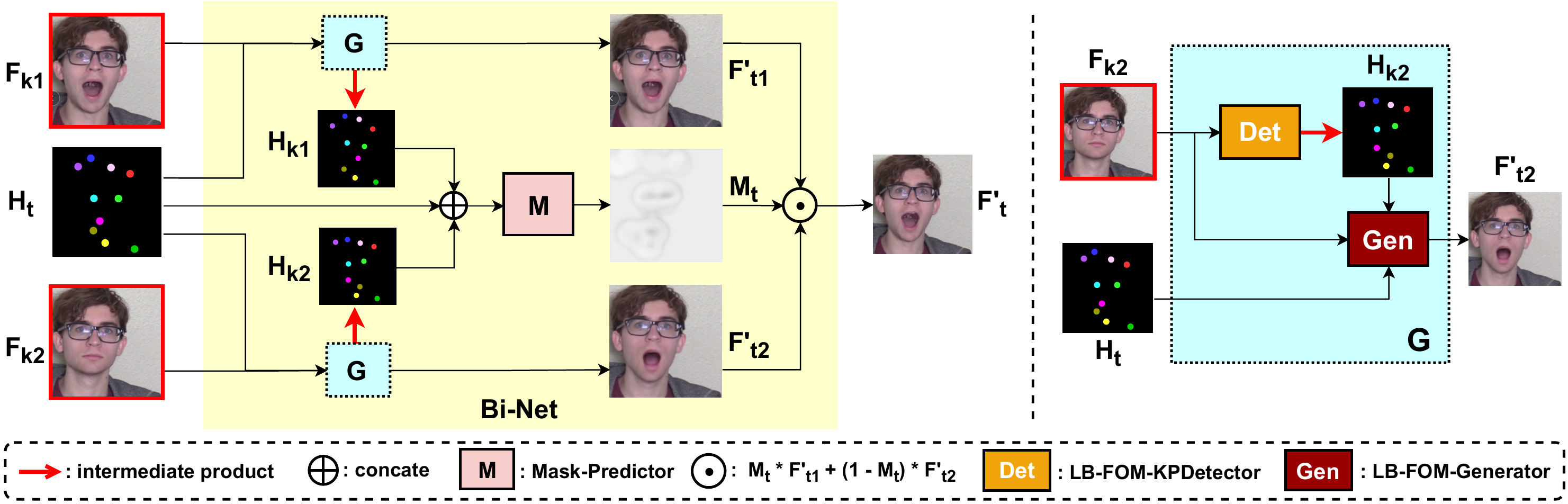}
  \caption{Architecture of the proposed Bi-Net. 
  According to two adjacent key frames ($F_{k1}$ and $F_{k2}$) and the keypoint heatmap ($H_t$) of the intermediate non-key frame ($F_t$), we can reconstruct $F_t$ via Bi-Net. 
  }
  \label{fig:Bi-Net}
\end{figure*}

\section{Method}
\label{sec:method}
\begin{figure}[htb]
  \centering
  \subfigure[Architecture of FOM\cite{fomm}.]{
      \includegraphics[width=1\linewidth]{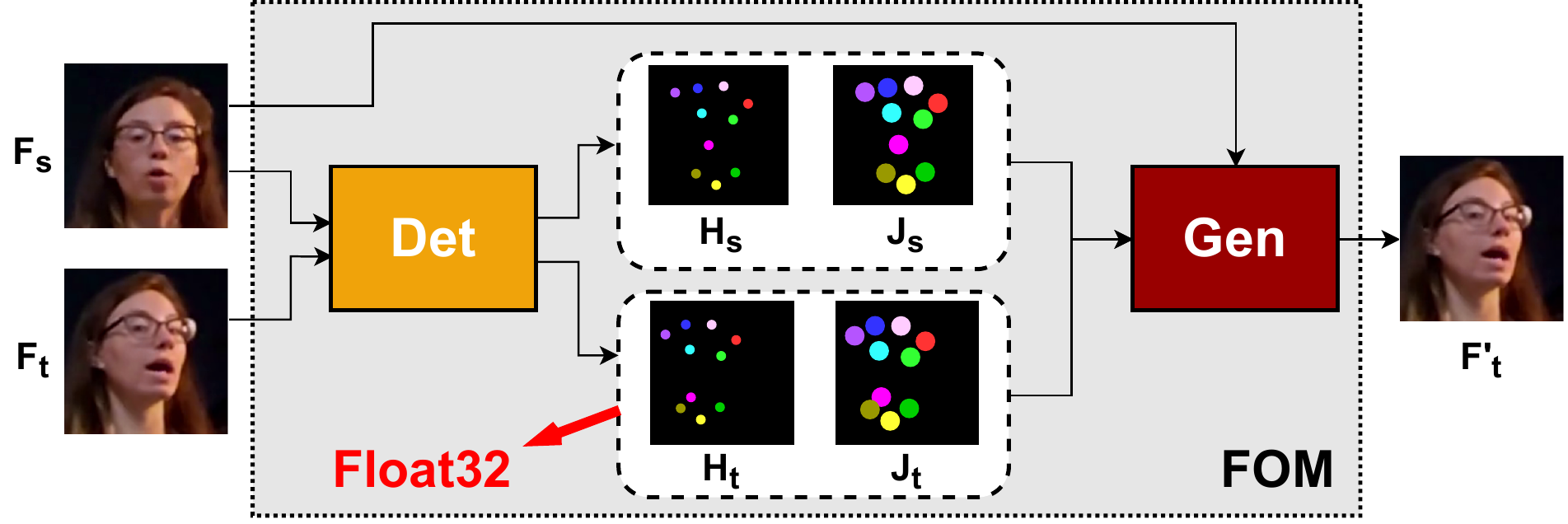}
      \label{fig:FOM}
  } \\
  \vspace{-0.2cm}
  \subfigure[Architecture of LB-FOM.]{
      \includegraphics[width=1\linewidth]{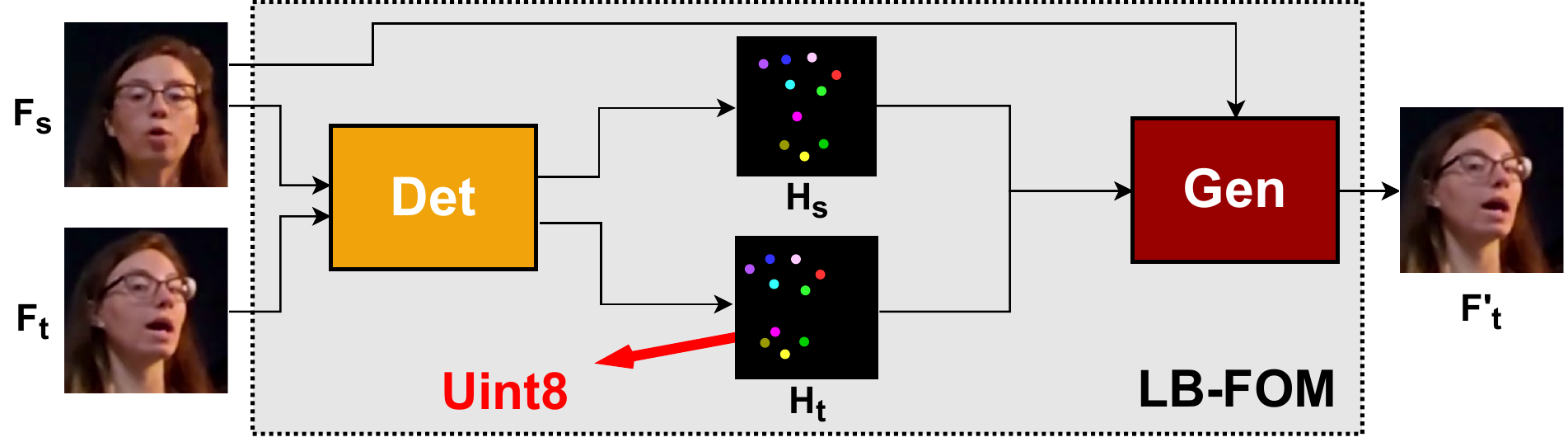}
      \label{fig:lb-fom}
  } \\
  \vspace{-0.2cm}
\caption{The architecture of FOM\cite{fomm} and LB-FOM.}
\vspace{-0.4cm}
\end{figure}

For a video sequence, at the sender, 
we sample the key frames according to a certain frequency 
(e.g., take one key frame every 5 frames), 
encode these key frames with 
VVenC\cite{vvenc}, an efficient VVC realization which has compression performance similar to VVC but with much less encoding complexity, 
and transmit the bitstream to the receiver. 
For non-key frames, we use the proposed LB-FOM-KPDetector in Figure~\ref{fig:lb-fom} to detect keypoints of type Uint8, encode them losslessly and transmit the bitstream to the receiver. 
At the receiver, the bitstream of VVenC is decoded to reconstruct the key frames. 
With the reconstructed key frames and the keypoints of the non-key frames, 
we utilize Bi-Net, the network of the proposed Pixel-wise Bi-Prediction method, to reconstruct the non-key frames. 
The network architecture is illustrated in Section ~\ref{sec:joint}.
Finally, we rearrange all frames in chronological order to get the final video streaming at the receiving end.
Details of the lossless coding scheme and bitrate calculation will be explained in Section ~\ref{sec:coding}.

\subsection{Bi-Net: Pixel-wise Bi-Prediction}
\label{sec:joint}
In Figure~\ref{fig:prediction_schemes}, Forward Prediction tends to lead to obvious jitter between the 5th and 6th frame, and Forward+Backward Prediction tends to lead to obvious jitter between the 3rd and 4th frame, damaging our subjective experience, which has been proved in Table ~\ref{tab:three_schemes}.
Thus, we propose a Pixel-wise Bi-Prediction method and design a Bi-Net to implement this idea.

As shown in Figure~\ref{fig:Bi-Net}, overall, the input of Bi-Net is two adjacent key frames ($F_{k1}$ and $F_{k2}$) and the keypoint heatmap of a non-key frame ($H_t$), and the output is the predicted non-key frame $F_t'$ corresponding to $H_t$.   
Specifically, we feed $F_{k1}$ and $H_t$ to G to get the intermediate product $H_{k1}$ and the reenacted result $F_{t1}'$. 
In the same way, $F_{k2}$ and $H_t$ are fed to G to get $H_{k2}$ and $F_{t2}'$.
Then, the concatenation of $H_{k1}$, $H_t$ and $H_{k2}$ is fed to the Mask-Predictor with an encoder-decoder architecture to predict a mask $M_t$ with value between 0 and 1 which acts on $F_{t1}'$ and $F_{t2}'$, obtaining the final result $F_{t}'$:
\begin{equation}
\label{equation:mask}
F_t' = M_t * F_{t1}' + (1 - M_t) * F_{t2}' .
\end{equation}

\noindent{The architecture of G is shown at the right of Figure~\ref{fig:Bi-Net}, and the details of LB-FOM is illustrated below.}

\begin{table}[tb]
\scriptsize
\renewcommand\arraystretch{1.15}
\begin{center}
\caption{The comparison of FOM\cite{fomm} and LB-FOM in bitrate and image synthesis quality.}
\label{tab:lb_fom}
    \begin{tabular}{|c|c|c|c|c|c|}
    \hline
    Method & Jac & dType & noise& Bytes/frame $\downarrow$ & PSNR(Y) $\uparrow$ \\
    \hline
    FOM \cite{fomm}  & Yes & Float32 & - & 240 & \textbf{36.76} \\
    FOM w/o Jac  & No & Float32 & - & \underline{80} & 36.28 \\
    LB-FOM w/o noise  & No & Uint8 & No & \textbf{20} & 36.06 \\
    LB-FOM  & No & Uint8 & Yes & \textbf{20} & \underline{36.47} \\
    \hline
    \end{tabular}
\end{center}
\vspace{-0.5cm}
\end{table}
\begin{figure*}[htb]
  \centering
  \includegraphics[width=1\linewidth]{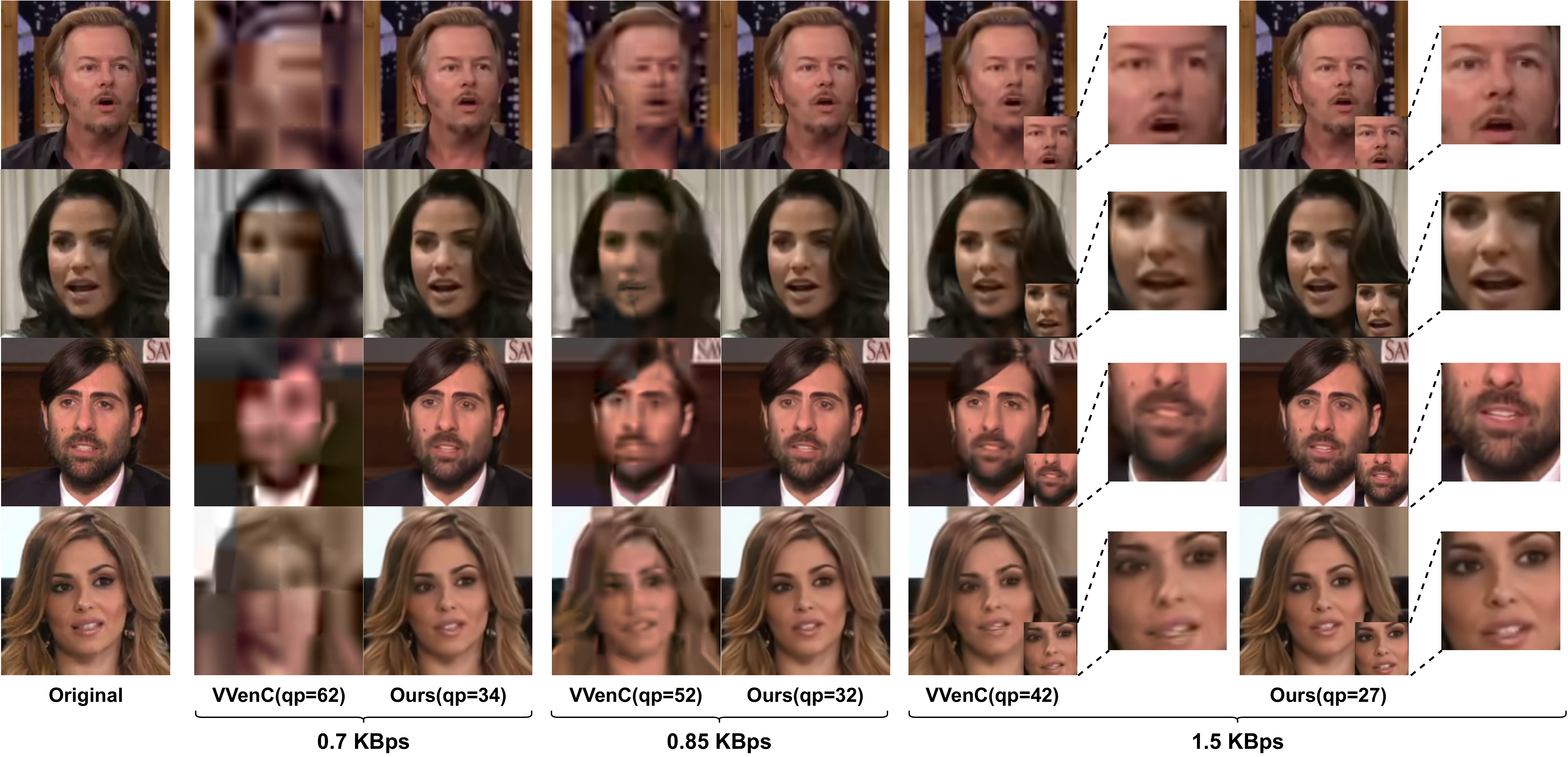}
  \caption{The reconstructed images using VVenC and our method under the same bitrate. 
  Mode: \textit{RA}, sampling-interval:10
  }
  \label{fig:result}
  \vspace{-0.3cm}
\end{figure*}

\vspace{2mm}
\noindent\textbf{LB-FOM}
Siarohin et al. proposed FOM\cite{fomm} to implement face animation.
As shown in Figure~\ref{fig:FOM}, FOM\cite{fomm} utilizes a KPDetector to predict the keypoints $H_s$ and corresponding Jacobian $J_s$ of the source image $F_s$,
and predict the keypoints $H_t$ and corresponding Jacobian $J_t$ of the target image $F_t$, where all data is of type Float32.
Then $H_s$, $J_s$, $H_t$, $J_t$ and the source image $F_s$ are fed to the generator to obtain the final result $F_t'$.

In FOM\cite{fomm}, each frame is represented by a fixed number of keypoints (e.g. 10) along with their Jacobians.
Each keypoint consists of an (x, y) coordinate, and the corresponding Jacobian is of size $2\times2$.
Since all data is of type Float32, for each frame, the keypoints are represented by 80 Bytes and the Jacobians are represented by 160 Bytes, resulting in a total bitrate of 240 Bytes/frame.

To lower the bitrate, we propose a LowBitrate-FOM (LB-FOM) in Figure~\ref{fig:lb-fom}.
Specifically, we remove the prediction of Jacobians and quantify the keypoints in [-1,1]  into range [0, 255], which means data of type Float32 is converted to type Uint8. 
In this way, the total bitrate decreases from 240 Bytes/frame to 20 Bytes/frame.
Table ~\ref{tab:lb_fom} shows the impact of such cuts.
In Table ~\ref{tab:lb_fom}, removing the Jacobian and converting Float32 to Uint8 directly lead to the continuous decline of PSNR, from 36.76 dB to 36.28 dB to 36.06 dB.
This is probably because direct quantization is non-differentiable, resulting in the failure of gradient back propagation.
When we add uniform noise to the predicted keypoints to simulate the quantization loss in the training phase, the PSNR value decreases only 0.29 dB compared to the original FOM\cite{fomm}, while the bitrate is reduced to one-twelfth of it, which is a good deal, demonstrating the effectiveness of our proposed LB-FOM.
An ablation study is conducted in Table ~\ref{tab:quantify} to further 
explore the issue about keypoints quantization.

\subsection{Training Schemes}
First, we take two frames from the same video at a time as paired training data and train LB-FOM in a supervised manner with (\ref{loss:total_loss}).
After training LB-FOM, fix the weights of it, that is, fix G, and train the Mask-Predictor to predict a valid mask $M_t$ which acts on $F_{t1}'$ and $F_{t2}'$ to synthesize the final $F_t'$ with (\ref{loss:total_loss}). 
Specifically, we take three frames from the same video at a time, and let the intermediate frame as the non-key frame and the rest two frames as the key frames.
After training the Mask-Predictor, finetune the whole Bi-Net for a few epochs.
Finally, quantify the keypoints directly rather than use uniform noise to simulate the quantization loss, and finetune another epoch to further improve the performance.

\vspace{0.15cm}
\noindent{\textbf{Loss functions.}}
We use L1, L2 and perceptual loss\cite{perceptual-loss} to optimize our model:
\begin{equation}
\label{loss:lb-fom l1}
L_{1} = \|F_{t}'-F_{t}\|_{1},
\end{equation}
\begin{equation}
\label{loss:lb-fom l2}
L_{2} = \|F_{t}'-F_{t}\|_{2},
\end{equation}
\begin{equation}
\label{loss:lb-fom perc}
L_{perc} = \sum_{i=1}^{n} \left\|\mathrm{VGG}_{i}\left(F_{t}\right)-\mathrm{VGG}_{i}(F_{t}')\right\|_{1},
\end{equation}
where $i$ is the selected layer indexes of VGG19. 
Full loss:
\begin{equation}
\label{loss:total_loss}
L_{full} = \lambda_{1}\cdot L_{1} + \lambda_{2}\cdot L_{2} + \lambda_{perc}\cdot L_{perc},
\end{equation}
where $\lambda_{1}$=$\lambda_{2}$=$\lambda_{perc}$=10.

\subsection{Lossless Coding Scheme and Bitrate Calculation}
\label{sec:coding}
In this paper, LB-FOM predicts 10 keypoints of type Uint8 for each non-key frame, so each non-key frame is represented by 20 Bytes if uncompressed. 
To encode these keypoints, we design a lossless coding scheme, 
in which way each non-key frame is compressed to 4.98 Bytes on average.
Specifically, we utilize intra and inter prediction, exponential Golomb coding and adaptive binary arithmetic coding\cite{binary-coding} to further reduce the bitrate.
More details can be found in $Sup.Mat.$.

The total bitstream is composed of two parts: VVenC bitstream from key frames and keypoint bitstream from non-key frames.
The sampled key frames are treated as a new video and VVenC is utilized to encode this video. The bitstream is determined by VVenC algorithm\cite{vvenc}.
As mentioned above, each non-key frame is represented by 4.98 Bytes on average with our lossless coding scheme. 
Assuming that the sampling interval is $N$, which means there is one key frame in every $N$ frames, we can calculate the average bitrate:
\begin{equation}
\label{equation:bitrate}
B_{avg} = [ B_k * 1 + B_{nk} * ( N - 1 ) ] / N ,
\end{equation}
where $B_k$ is the bitrate of the key frame,  $B_{nk}$ is the bitrate of the non-key frame, and $B_{avg}$ is the average bitrate.

\section{Experiments}
\subsection{Implementation Details}
\noindent{\textbf{Datasets.}}
In this paper, we use DFDC\cite{dfdc} as our trainset. 
DFDC dataset consists of over 15,000 face videos and each video consists of nearly 300 frames.
To test our method, we download 15 videos from YouTube in VoxCeleb2\cite{voxceleb2} as our testset, and each video consists of nearly 300 frames.
All experiments were conducted at 256$\times$256 image resolution, while 
Table ~\ref{tab:full_data} further shows some experimental results at 512$\times$512 resolution.
For all videos, frame rate is 30 FPS.

\vspace{0.1cm}
\noindent{\textbf{Metrics.}}
We use PSNR(Y) (in dB), the Peak Signal-to-Noise Ratio value of Y channel in yuv format, as the main objective metric.
For subjective metrics, in Table ~\ref{tab:vvenc_ours}, \textit{fidelity} score reflects the degree of detail retention and \textit{aesthetics} score reflects the overall video visual quality.
Specifically, we simultaneously show two videos encoded by VVenC and our method respectively and tell the users to choose the better one in fidelity or aesthetics, so the score is actually the user preference percentage.
In Table ~\ref{tab:three_schemes} and Table ~\ref{tab:quantify}, 
the \textit{smoothness} score reflects the degree of video smoothness and fluency. 
These methods are compared in pairs. For each method, \textit{smoothness}-score = the number of times it was preferred / the total number of times it was compared.

\begin{table}[t]
\scriptsize
\renewcommand\arraystretch{1.15}
\begin{center}
\caption{
The quantitative results corresponding to Figure~\ref{fig:result}.
bpp: bits-per-pixel, KB/s: 1024-Bytes-per-second.
"Ours-QP34-N10" means the sampling interval is 10 and for key frames encoded by VVenC, the QP value is 34.
The corresponding compressed videos can be seen in $Sup.Mat.$.
}
\label{tab:vvenc_ours}
    \begin{tabular}{|c|c|c|c|c|c|}
    \hline
    \multirow{2}{*}{Method} & \multicolumn{2}{c|}{Bitrate}
    & \multirow{2}{*}{PSNR(Y)} & \multicolumn{2}{c|}{User Study}\\
    \cline{2-3}
    \cline{5-6}
    &  $10^{-3}$bpp & KB/s &  & Fidelity & Aesthetics \\
    \hline
    VVenC-QP62 &  2.97 & 0.71 & 24.13 & 0.00 & 0.00 \\
    Ours-QP34-N10 & 2.96 & 0.71 & \textbf{34.42} & \textbf{1.00} & \textbf{1.00} \\
\hline
    VVenC-QP52 &  3.76 & 0.90 & 28.93 & 0.00 & 0.00 \\
    Ours-QP32-N10 &  3.55 & 0.85 & \textbf{35.04} & \textbf{1.00} & \textbf{1.00} \\
    \hline
    VVenC-QP42 & 6.21 & 1.49 & 34.47 & 0.15 & 0.00 \\
    Ours-QP27-N10 &  6.12 & 1.47 & \textbf{36.23} & \textbf{0.85} & \textbf{1.00} \\
    \hline
    VVenC-QP42 &  6.21 & 1.49 & 34.47 & 0.16 & 0.18 \\
    Ours-QP34-N10 &  \textbf{2.96} & \textbf{0.71} & 34.42 & \textbf{0.84} & \textbf{0.82}\\
    \hline
    \end{tabular}
\end{center}
\vspace{-0.3cm}
   \end{table}

\begin{figure}[t]
  \centering
  \includegraphics[width=1\linewidth]{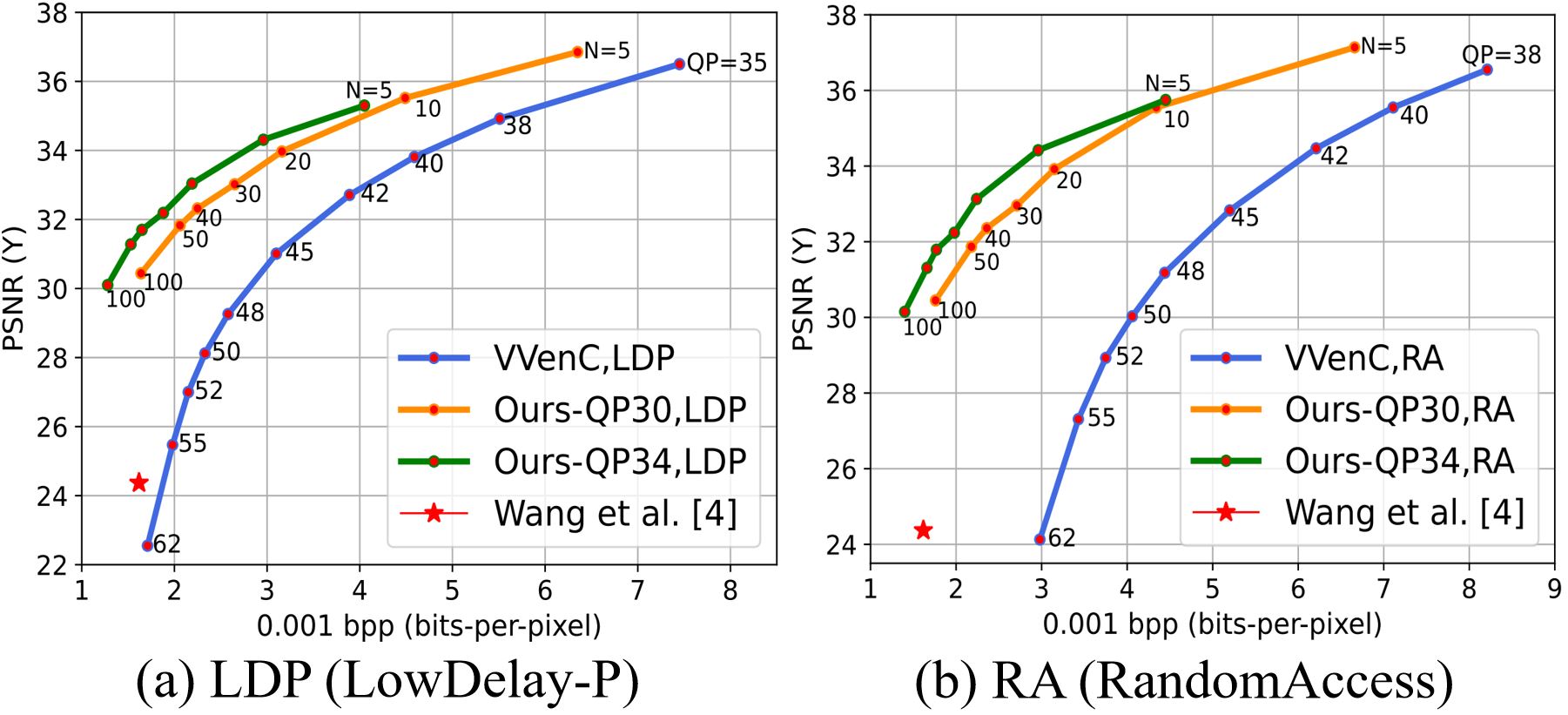}
  \caption{The performance comparison of VVenC and our method in both \textit{LDP} and \textit{RA} mode.
  \textit{LDP}: BD-rate = \textbf{-43.12\%}, \textit{RA}: BD-rate = \textbf{-58.48\%}.
  }
  \label{fig:zhexian}
\end{figure}

\begin{table}[t]
\footnotesize
\renewcommand\arraystretch{1.15}
\begin{center}
\caption{
More results at 256$\times$256 and 512$\times$512 resolutions.
Mode: \textit{RA}. KB/s: 1024-Bytes-per-second.
"Ours-QP34-N30" means the sampling interval is 30 and for key frames encoded by VVenC, the QP value is 34.
}
\label{tab:full_data}
    \begin{tabular}{|c|c|c|c|c|}
    \hline
    \multirow{2}{*}{Method} 
    & \multicolumn{2}{c|}{256$\times$256} 
    & \multicolumn{2}{c|}{512$\times$512} 
    \\
    \cline{2-3}
    \cline{4-5}
      & KB/s & PSNR(Y) & KB/s & PSNR(Y)\\
    \hline
    VVenC-QP62 & 0.71 & 24.13 & 0.82 & 26.82\\
    VVenC-QP52 & 0.90 & 28.93 & 1.23 & 31.92  \\
    \hline
    Ours-QP34-N30 & 0.48 & 32.24 &  0.86 & 31.59 \\
    Ours-QP36-N30 & 0.42 & 31.87 &  0.74 & 31.49 \\
    Ours-QP36-N10 & 0.60 & 33.81 &  1.13 & 33.47 \\
    \hline
    \end{tabular}
\end{center}
   \end{table}

\begin{table}[t]
\scriptsize
\renewcommand\arraystretch{1.15}
\begin{center}
\caption{An ablation study on three prediction methods.}
\label{tab:three_schemes}
\setlength\tabcolsep{8pt}
    \begin{tabular}{|c|c|c|c|}
    \hline
    Method & Forward  & Forward+Backward  & Pixel-wise Bi-Prediction  \\
    \hline
    PSNR(Y)  & 35.45 & \underline{36.05} & \textbf{36.47} \\
    Smoothness & 0.021 & \underline{0.083} & \textbf{0.937} \\
    \hline
    \end{tabular}
\end{center}
\end{table}

\subsection{Comparison Results}
We mainly compare our proposed method with VVenC in Figure~\ref{fig:result} and Figure~\ref{fig:zhexian}.
As shown in Figure~\ref{fig:result}, at the same bitrate, the visual quality of our generated images are obviously better than that of VVenC.
For example, at the bitrate of 0.7 KB/s, VVenC obtained the performance of QP=62, while in our method, we encoded the key frames at QP=34 and generated the non-key frames almost with the same quality as the key frames, that is, cost the bitrate of QP=62 while achieved the quality of QP=34.
Specific values of bitrate and PSNR are shown in Table ~\ref{tab:vvenc_ours}, where the results demonstrate that our method achieves obviously better performance than VVenC in detail retention and overall quality.
Additionally, we show more results at 256$\times$256 and 512$\times$512 resolutions in Table ~\ref{tab:full_data}.
Complete experimental data is shown in $Sup.Mat.$.

Figure~\ref{fig:zhexian} shows the quantitative results in \textit{LDP} mode and \textit{RA} mode. Without the loss of generality, we take Figure~\ref{fig:zhexian}(a) as example.
We choose VVenC as our baseline, and test its performance from QP=35 to QP=62.
Higher QP value means higher compression ratio.
As mentioned above, our method can implement dynamically-adjustable bitrate through two channels: adjust the sampling interval of key frames or adjust the compression ratio (that is, QP value) of key frames.
In Figure~\ref{fig:zhexian}(a), the orange broken line 
shows the performance of different sampling intervals from 5 to 100 under QP=30,
and the green broken line is the performance under QP=34. 
In fact, QP value and the sampling interval can be determined freely, not limited to those listed.
In addition, we also compared our method with Wang et al.\cite{nvidia-maxine}.
Since there is no released test code, we applied the data released in their paper and drew it in Figure~\ref{fig:zhexian}.
It can be found that our method achieves significantly better performance than VVenC and Wang et al.\cite{nvidia-maxine} in either \textit{LDP} mode or \textit{RA} mode.

\subsection{Ablation Studies}
\label{sec:ablation}
\begin{figure}[tbp]
  \centering
  \includegraphics[width=1\linewidth]{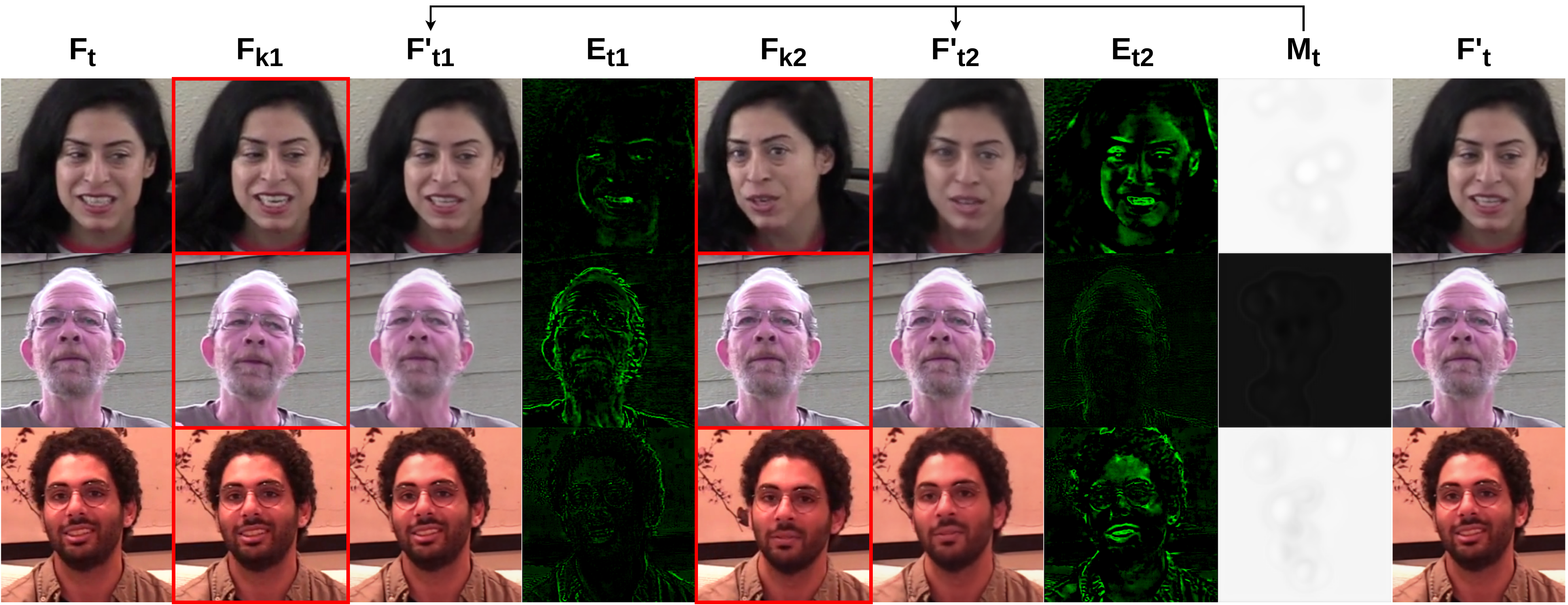}
  \caption{The intermediate products generated in the inference process.
  $F_t$ is the target non-key frame to be generated.
  $F_{k1}$ \& $F_{k2}$ are two adjacent key frames, and
  $F_{t1}'$ \& $F_{t2}'$ are their predictive results using G, respectively.
  $E_{t1}$ is the error map between $F_{t}$ and $F_{t1}'$, and 
  $E_{t2}$ is the error map between $F_{t}$ and $F_{t2}'$.
  $M_t$ is the predicted mask which operates on $F_{t1}'$ and $F_{t2}'$.
  $F_t'$ is the final synthesized result.
  }
  \label{fig:mask}
  \vspace{-0.3cm}
\end{figure}

As mentioned in Section ~\ref{sec:joint}, Forward Prediction and Forward+Backward Prediction have the disadvantage of generating unsmooth video, so we propose the Pixel-wise Bi-Prediction method.
To demonstrate its effectiveness, we compare these three prediction methods, and show results in Table ~\ref{tab:three_schemes}, where the Pixel-wise Bi-Prediction method achieves the highest PSNR(Y) score and has the best performance in video smoothness.
In Figure~\ref{fig:mask}, we visualize the intermediate products generated in the inference process.
It can be found that the predicted mask $M_t$ tends to trust the frame predicted by the key frame that looks more like $F_t$, which makes sense.
For example, in the first row, $F_{k1}$ looks more like $F_t$ than $F_{k2}$, so the predicted mask gives more weight to $F_{t1}'$ than $F_{t2}'$.

We also conduct an ablation study in Table ~\ref{tab:quantify} to demonstrate the necessity of quantifying keypoints during training and the necessity of adding uniform noise to simulate the quantization loss. 
In Table ~\ref{tab:quantify}, quantifying keypoints only in the testing phase results in relatively bad objective performance.
Quantifying keypoints in the training phase without adding noise leads to an obvious decline in \textit{smoothness} score.  
By contrast, quantifying keypoints in the training phase and adding uniform noise to simulate the quantization loss achieves satisfactory performance in both objective and subjective experiments.

\begin{table}[tp]
\footnotesize
\renewcommand\arraystretch{1.15}
\begin{center}
\caption{An ablation study on the quantization of keypoints.
}
\label{tab:quantify}
    \begin{tabular}{|c|c|c|c|}
    \hline
    Quantify in training & add noise & PSNR(Y) & Smoothness \\
    \hline
    No  & - & 36.04  & \textbf{0.85} \\
    Yes  & No & \underline{36.06}  & 0.81 \\
    Yes  & Yes & \textbf{36.47} & \underline{0.83}\\
    \hline
    \end{tabular}
\end{center}
\vspace{-0.5cm}
\end{table}

\section{Conclusion}
In this paper, we propose a hybrid compression scheme for face video, 
which combines the pixel-level precise recovery capability of traditional coding with the generation capability of deep learning based on abridged information,
achieving the purpose of high quality face video synthesis at ultra-low bitrate.
Our proposed Pixel-wise Bi-Prediction method helps generate smooth videos, greatly improving our subjective experience. 
Without introducing any additional bitrate, a series of completely fair experiments have demonstrated the effectiveness of our proposed scheme. 
Moreover, this scheme has the characteristic of dynamically-adjustable bitrate and adapts to any existing encoder/configuration to deal with different encoding requirements, resulting in a strong practicability.
Our future work includes better keypoint compression method and delay optimization.

\section*{Acknowledgements}

This work was supported in part by MoE-China Mobile Research Fund Project under Grant MCM20180702, the National Key Research and Development Project of China under Grant 2019YFB1802701, Chinese National Science Funding 62132006, and Shanghai Key Laboratory of Digital Media Processing and Transmissions.

\bibliographystyle{ieeebib}
\bibliography{my}

\appendix
\clearpage

In this supplementary, we provide more details and results to make our paper more comprehensive. 

\section{Related Work}
\subsection{Traditional Video Compression}
In the past few years, many traditional video compression methods have been proposed, such as HEVC\cite{hevc} and VVC\cite{vvc}. Based on hybrid video compression framework, traditional methods utilize intra and inter prediction to remove spatial and temporal redundancy by some manually designed schemes. 
VVC\cite{vvc} represents the most advanced traditional method. Compared with HEVC\cite{hevc}, VVC can save about 50\% of the bitrate while maintaining the similar visual quality.  
Fraunhofer HHI further released an efficient VVC realization VVenC\cite{vvenc} which has compression performance similar to VVC but with much less encoding complexity.

\subsection{Deep Talking Head Approaches}
Wang et al. proposed the vid2vid\cite{wang2018videotovideo,wang2019fewshot}
methods to implement talking head synthesis, 
but these methods need the transmission of edge maps, which have a relatively high bandwidth cost.
The face-reenactment methods\cite{nirkin2019fsgan,bilayer} also can be applied to implement talking head synthesis. However, the bandwidth cost of the required representations is really high.
There are also some 3D-based methods\cite{3d-avatars,kim2018dvp} to implement portrait synthesis like DVP\cite{kim2018dvp}. However, this method is subject-specific and needs a thousand images per subject for training.
First order model(FOM)\cite{fomm} uses a representation of a set of learned keypoints along with their local affine transformations to implement the face animation and achieves satisfactory performance especially when the source actor and the target actor share the same identity. 
Also, the bandwidth cost of FOM\cite{fomm} is relatively low compared to other methods\cite{wang2018videotovideo,wang2019fewshot,nirkin2019fsgan,bilayer}.
Based on this, in our method, we designed a low bitrate version of FOM\cite{fomm}: LowBitrate-FOM (LB-FOM).

\section{Lossless Coding Scheme}
\label{sec:lossless_coding}
\label{sec:coding}

\begin{figure}[tb]
\centering
\subfigure[Intra Prediction.]{
\label{intra}
\includegraphics[width=0.38\linewidth]{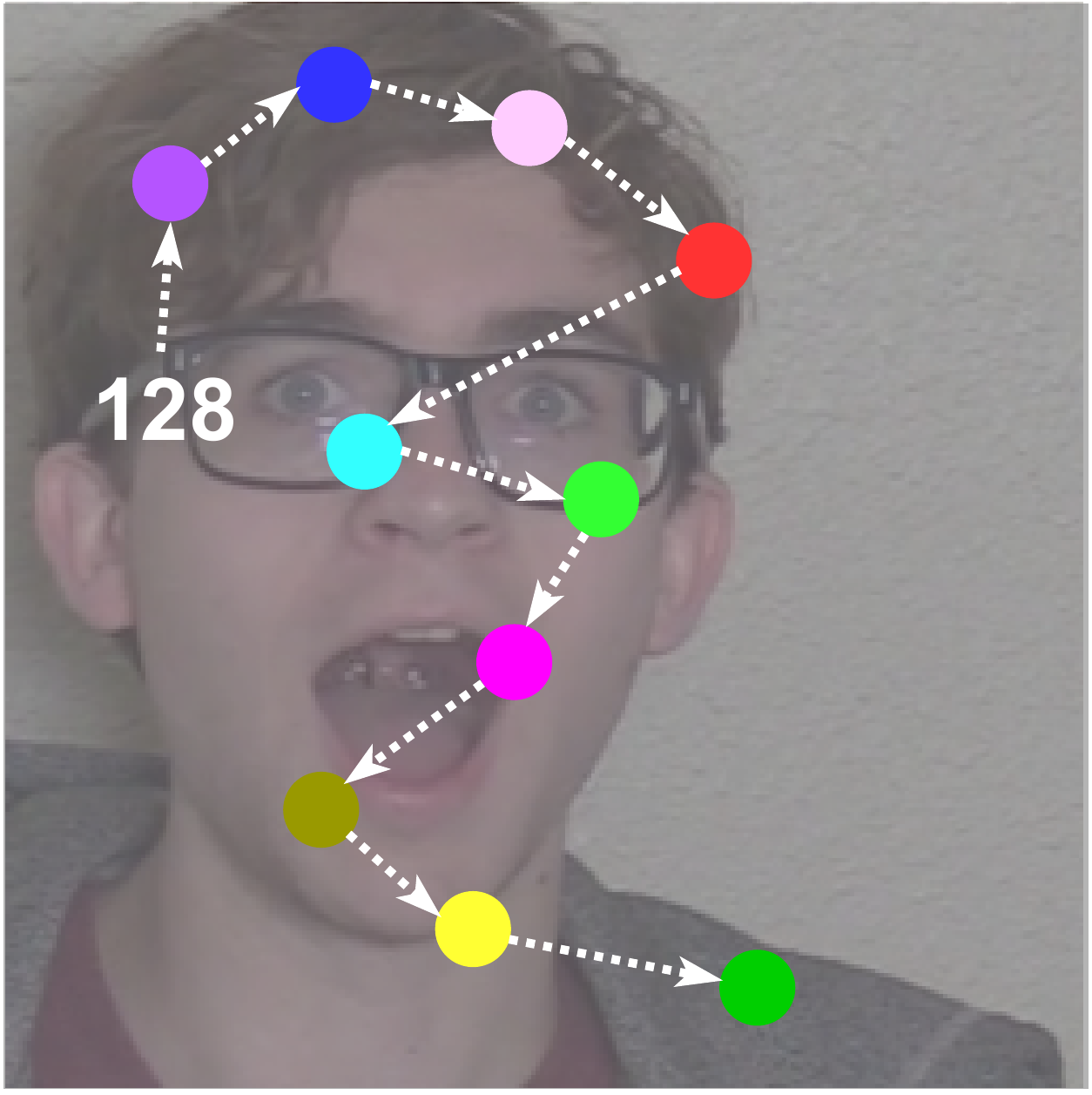}}
\hspace{0.1cm}
\subfigure[Inter Prediction.]{
\label{inter}
\includegraphics[width=0.47\linewidth]{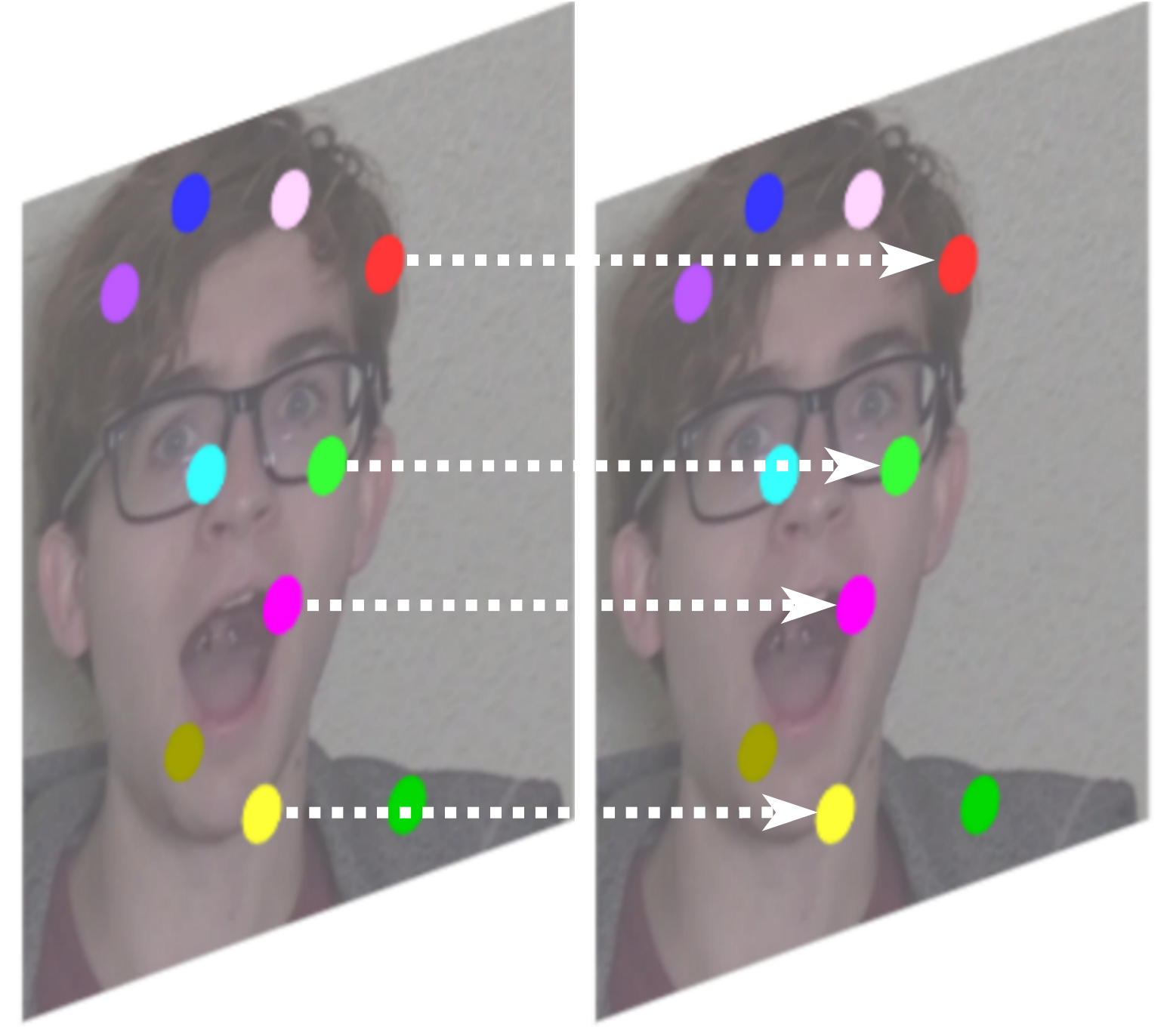}}
\caption{Intra \& Inter Prediction in our lossless coding scheme.
Intra Prediction: use $(N-1)$th point to predict $N$th point. Inter Prediction: use $(K-1)$th frame to predict $K$th frame.}
\label{prediction}
\end{figure}

In this paper, we let LB-FOM to predict 10 keypoints for each frame. Each keypoint consists of one x coordinate and one y coordinate which is of type Uint8, which means each non-key frame is represented by 20 Bytes if uncompressed. To further compress these values, we design a novel method which reduces the bitrate to 4.98 Bytes per non-key frame. 
The compression ratio of each step is shown in Table ~\ref{tab:compress-ratio}. 

\vspace{2mm}
\noindent{\textbf{Intra and Inter Prediction}}
For video sequence, the adjacent frames have high similarity, which may cause information redundancy. We utilize inter prediction to remove the temporal redundancy. As shown in Figure \ref{inter}, we use keypoints in the $(K-1)$th frame to predict the corresponding keypoints in the $K$th frame. That means we do not straightly encode the keypoint value of the $K$th frame, but encode the residual keypoint value of the $(K-1)$th frame and the $K$th frame. 
For the first non-key frame, we apply intra prediction to remove spatial redundancy. As shown in Figure \ref{intra}, we use the $(N-1)$th keypoint to predict the $N$th keypoint. And for the first keypoint in the first non-key frame, we use 128 which is the middle value of coordinate to estimate it. Combined with inter and intra prediction, the bitrate is reduced to 11.7 Bytes per non-key frame.

\vspace{2mm}
\noindent{\textbf{Variable-Length Coding}}
As a Variable-Length Coding algorithm, exponential Golomb coding benefits more when small numbers appear frequently. Duo to keypoints change slightly between adjacent frames, the residual value is small which fits perfectly with the characteristic of exponential Golomb coding. Thus we use zero-order exponential Golomb coding to encode the residual value. 

What's more, we use adaptive binary arithmetic coding (an entropy coding)\cite{binary-coding} to further reduce the bitrate. At last, the bitrate is reduced to 4.98 Bytes per non-key frame.

\section{Delay Analysis}
In this part, we will discuss the delay problem.
Suppose the sampling interval is $N$.
When adopting \textit{LDP} (LowDelay-P) mode of VVenC, the delay is ($N$ - 1) frames at the receiver compared to the sender.
When adopting \textit{RA} (RandomAccess) mode of VVenC and GOP-size is set to $k$, the delay is ($k$*$N$ - 1) frames at the receiver compared to the sender.
Consequently, for video conferencing and other scenes with lowdelay requirements, we encode key frames in \textit{LDP} mode. 
For scenarios that do not have lowdelay requirements such as offline video storage, \textit{RA} mode is also an option.

\begin{table}[tbp]
\renewcommand\arraystretch{1.15}
\scriptsize
  \centering
  \caption{The compression ratio of each step in our scheme.
  1st: remove Jacobian. 
  2nd: convert data type from Float32 to Uint8.  
  3rd: Intra \& Inter Prediction.  
  4th: Variable-Length Coding \& Arithmetic Coding.
  Bpf: Bytes-per-frame.}
  \setlength\tabcolsep{9pt}
  \vspace{0.2cm}
    \begin{tabular}{|c|c|c|c|c|c|}
    \hline
    & Origin &  1st  & 2nd  & 3rd  & 4th   \\
    \hline
    CompressRatio    &  -   &  0.333   &  0.250  & 0.585  & 0.425  \\
    \hline
    non-key-frame / Bpf    & 240   & 80    & 20    &  11.70    &   4.98 \\
    \hline
    \end{tabular}%
  \label{tab:compress-ratio}%
  \vspace{-0.3cm}
\end{table}%

\section{Experimental Setups}
During training, we apply Adam optimizer~\cite{Kingma2015AdamAM} with the multi-step learning rate, $\beta_{1}=0.5$ and $\beta_{2}=0.999$. The project is implemented by PyTorch and our training is carried out on 3 NVIDIA TITAN Xp GPUs with batch size 12.

\vspace{2mm}
\noindent\textbf{VVenC setups.}
In \textit{LDP} mode, GOPSize = 8 and IntraPeriod = -1.
In \textit{RA} mode, GOPSize = IntraPeriod = 32.

\section{Complete experimental data}
In RA mode, the complete experimental data of VVenC\cite{vvenc} and our method is shown in Table ~\ref{tab:vvenc_ra} and Table ~\ref{tab:ours_ra}.
And in LDP mode, the complete experimental data of VVenC and our method is shown in Table ~\ref{tab:vvenc_ldp} and Table ~\ref{tab:ours_ldp}.
In Table ~\ref{tab:vvenc_ra} and Table ~\ref{tab:vvenc_ldp}, we show the bitrate in three units:
bpf (bits-per-frame), bpp (bits-per-pixel) and KBps (1024-Bytes-per-second).

In Table ~\ref{tab:ours_ra} and Table ~\ref{tab:ours_ldp}, 
we show the experimental results of our method under different QP values and sampling intervals.
For a certain sampling interval, we can divide all frames into two categories: key frames and non-key frames, and the total bitrate is composed of key-frame bitrate and non-key-frame bitrate.
From Section ~\ref{sec:coding}, the bitrate of non-key frames is 20 Bpf (Bytes-per-frame) without compression. 
With our lossless coding scheme described in Section ~\ref{sec:coding}, we can compress the non-key-frame bitrate from 20 Bpf to 4.98 Bpf on average.
Note that the specific bitrate of non-key frames after compression is slightly different
under different sampling intervals.
The key frames are seen as a video and are encoded by VVenC with a predefined QP value,
so the bitrate of key frames is different when adopting different QP values to compress the key frames. 
Under a certain sampling interval and QP value, we can calculate the average bitrate according to the corresponding key-frame bitrate and non-key-frame bitrate. 
In Table ~\ref{tab:ours_ra} and Table ~\ref{tab:ours_ldp}, we show the bitrate in four units:
Bpf (Bytes-per-frame), bpf (bits-per-frame), bpp (bits-per-pixel) and KBps (1024-Bytes-per-second).

From the comparison between Table ~\ref{tab:vvenc_ra} and Table ~\ref{tab:ours_ra} and
the comparison between Table ~\ref{tab:vvenc_ldp} and Table ~\ref{tab:ours_ldp},
it can be found that compared to VVenC, our method achieves higher PSNR value under the same bitrate 
and costs lower bitrate when achieving the same PSNR value.

\begin{table*}[htp]
\renewcommand\arraystretch{1.15}
\begin{center}
\caption{Experimental results of VVenC under various QP values. Mode: RA.}
\label{tab:vvenc_ra}
\setlength\tabcolsep{9pt}
\vspace{0.2cm}
    \begin{tabular}{|c|c|c|c|c|c|c|c|c|c|}
    \hline
    QP & 32 & 33 & 34 & 35 & 36 & 37 & 38 & 39 & 40  \\
    \hline
    bpf & 903 & 817	& 747 & 686	& 631 & 582 & 538 & 502 & 466 \\
    $10^{-3}$bpp &  13.78 & 12.47 & 11.40 & 10.47 & 9.63 & 8.88 & 8.21 & 7.66 & 7.11  \\
    KBps & 3.31	&2.99&	2.74&	2.51&	2.31&	2.13	&1.97&	1.84&	1.71  \\
    PSNR(Y) & 39.29 & 38.81 & 38.40 & 37.97 & 37.51 & 37.03 & 36.55 & 36.05 & 35.55  \\
    \hline
    QP & 42 & 45 & 48 & 50 & 52 & 55 & 58 & 60 & 62  \\
    \hline
    bpf & 407 & 341 & 291 & 266 & 246 & 225 & 206 & 200 & 195  \\
    $10^{-3}$bpp & 6.21 & 5.2 & 4.44 & 4.06 & 3.75 & 3.43 & 3.14 & 3.05 & 2.98  \\
    KBps & 1.49&	1.25&	1.07&	0.97&	0.9&	0.82&	0.75&	0.73&	0.71   \\
    PSNR(Y) & 34.47 & 32.83 & 31.18 & 30.03 & 28.93 & 27.31 & 25.80 & 24.91 & 24.13  \\
    \hline
    \end{tabular}
\end{center}
\end{table*}
\begin{table*}[htp]
\small
\renewcommand\arraystretch{1.15}
\vspace{-0.4cm}
\begin{center}
\caption{Experimental results of our method under different QP values and sampling intervals. Mode: RA.}
\label{tab:ours_ra}
\setlength\tabcolsep{8pt}
\vspace{0.2cm}
    \begin{tabular}{|c|c|c|c|c|c|c|c|c|c|c|c|}
    \hline
    QP & \multicolumn{2}{c|}{sampling-interval} & 5 & 10 & 20 & 30 & 40 & 50 & 60 & 80 & 100  \\
    \hline
    & \multicolumn{2}{c|}{non-key-frame / Bpf}  & 5.358 & 5.079 & 4.964 & 4.922 & 4.907 & 4.985 & 4.887 & 4.876 & 4.873  \\
    \hline
    \multirow{5}{*}{27} & \multicolumn{2}{c|}{key-frame / bpf} & 2949& 3648 & 4822& 5859  & 6448 & 7173 & 7984 & 9149 & 10315 \\
    \cline{2-12}
    & \multirow{3}{*}{average-bitrate} & bpf & 624.1 & 401.4 & 278.8 & 233.4 & 199.5 & 181.8 & 171.5 & 152.9 & 141.7  \\
    & & $10^{-3}$bpp & 9.52 & 6.12 & 4.25 & 3.56 & 3.04 & 2.77 & 2.62 & 2.33 & 2.16  \\
    & & KBps & 2.29 & 1.47 & 1.02 & 0.85 & 0.73 & 0.67 & 0.63 & 0.56 & 0.52  \\
    \cline{2-12}
    & \multicolumn{2}{c|}{PSNR(Y)} & 37.98  & 36.23 & 34.43 & 33.33 & 32.70 & 32.12 & 31.49 & 31.06 & 30.58  \\
    \hline

    \multirow{5}{*}{30} & \multicolumn{2}{c|}{key-frame / bpf} & 2010 & 2478 & 3373	& 4182 & 4663 & 5214 & 5876 & 6814 & 7704\\
    \cline{2-12}
    &  \multirow{3}{*}{average-bitrate} & bpf & 436.3	& 284.4	& 	206.4	& 	177.5	& 	154.8	& 	142.7	& 	136.4	& 	123.7	& 	115.6 \\
    & &$10^{-3}$bpp & 6.66 & 4.34 & 	3.15& 	2.71& 	2.36& 	2.18& 	2.08& 	1.89& 	1.76 \\
    & & KBps & 1.6  & 1.04 &	0.76  &	0.65 &	0.57 &	0.52 &	0.5 &	0.45 &	0.42\\
    \cline{2-12}
    & \multicolumn{2}{c|}{PSNR(Y)} & 37.14& 	35.55& 	33.92& 	32.96& 	32.36& 	31.87& 	31.24& 	30.91& 	30.45   \\
    \hline
    
    \multirow{5}{*}{32} & \multicolumn{2}{c|}{key-frame / bpf} & 1593 &	1963 &	2687 &	3363 &	3783 &	4229 &	4827 &	5599 &	6337  \\
    \cline{2-12}
    &  \multirow{3}{*}{average-bitrate} &bpf &  352.9 & 	232.9 & 172.1 & 	150.2 & 	132.8	 & 123	 & 118.9 & 	108.5 & 	102  \\
    & & $10^{-3}$bpp & 5.38 &	3.55 &	2.63 &	2.29 &	2.03 &	1.88 &	1.81	 &1.66 &	1.56  \\
    & & KBps &1.29  & 0.85	 &0.63	 &0.55   &	0.49 &	0.45  &	0.44     &	0.4	 &0.37  \\
    \cline{2-12}
    & \multicolumn{2}{c|}{PSNR(Y)} & 36.47 &	35.04 &	33.58 &	32.63 &	32.10	 &31.60 &	31.07 &	30.65 &30.30   \\
    \hline
    
    \multirow{5}{*}{34} & \multicolumn{2}{c|}{key-frame / bpf} & 1286 &	1573 &	2182 &	2752 &	3119 &	3520 &	3991 &	4648 &	5308\\
    \cline{2-12}
    &  \multirow{3}{*}{average-bitrate} & bpf & 291.5 &	193.9 &	146.8 &	129.8 &	116.2 &	108.8 &105 &	96.6 &	91.7  \\
    & & $10^{-3}$bpp &4.45 &	2.96 &	2.24 &	1.98 &	1.77 &	1.66 &	1.6	 &1.47 &1.4  \\
    & & KBps & 1.07  &  0.71  &	0.54 &	0.48 &	0.43 &	0.4 &	0.38 &	0.35 &	0.34  \\
    \cline{2-12}
    & \multicolumn{2}{c|}{PSNR(Y)} & 35.75& 	34.42& 	33.13& 	32.24& 	31.79& 	31.31& 	30.85& 	30.48& 	30.15   \\
    \hline
    
    \multirow{5}{*}{36} & \multicolumn{2}{c|}{key-frame / bpf} & 1053&	1280&	1787&	2276&	2574&	2930&	3347&	3891&	4455 \\
    \cline{2-12}
    &  \multirow{3}{*}{average-bitrate} & bpf & 244.9	&164.6&	127.1&	113.9&	102.6&	97&	94.2&	87.2&	83.1 \\
    & & $10^{-3}$bpp & 3.74	&2.51&	1.94&	1.74&	1.57&	1.48&	1.44&	1.33&	1.27   \\
    & & KBps & 0.9   &	0.6	&0.47 &	0.42&	0.38&	0.36 &	0.34&	0.32&	0.3  \\
    \cline{2-12}
    & \multicolumn{2}{c|}{PSNR(Y)} & 34.99&	33.81&	32.65&	31.87&	31.41&	30.98&	30.51&	30.21&	29.89   \\
    \hline
    
    \multirow{5}{*}{38} & \multicolumn{2}{c|}{key-frame / bpf} & 868	&1048&	1476&	1864&	2135&	2420&	2779&	3264&	3740\\
    \cline{2-12}
    &  \multirow{3}{*}{average-bitrate} & bpf & 207.9&	141.4	&111.5&	100.2&	91.6&	86.8&	84.3&	79.3&	76   \\
    & & $10^{-3}$bpp & 3.17	&2.16&	1.7	&1.53&	1.4&	1.32&	1.29&	1.21&	1.16  \\
    & & KBps & 0.76       &	0.52&	0.41 &	0.37&	0.34&	0.32&	0.31&	0.29&	0.28 \\
    \cline{2-12}
    & \multicolumn{2}{c|}{PSNR(Y)} & 34.14&	33.05&	32.01&	31.26&	30.85&	30.54&	30.10&	29.79	&29.53 \\
    \hline
    \end{tabular}
\end{center}
\end{table*}
\begin{table*}[htp]
\renewcommand\arraystretch{1.15}
\begin{center}
\caption{Experimental results of VVenC under various QP values. Mode: LDP.}
\label{tab:vvenc_ldp}
\setlength\tabcolsep{9pt}
\vspace{0.2cm}
    \begin{tabular}{|c|c|c|c|c|c|c|c|c|c|}
    \hline
    QP & 32 & 33 & 34 & 35 & 36 & 37 & 38 & 39 & 40  \\
    \hline
    bpf & 679 & 605 & 542 & 488 & 439 & 398 & 361 & 330 & 301  \\
    $10^{-3}$bpp & 10.36 & 9.23 & 8.27 & 7.45 & 6.7 & 6.07 & 5.51 & 5.04 & 4.59  \\
    KBps & 2.49	&2.22&	1.98&	1.79&	1.61&	1.46&	1.32&	1.21&	1.1  \\
    PSNR(Y) & 38.00 & 37.51 & 37.01 & 36.50 & 35.97 & 35.45 & 34.92 & 34.36 & 33.81  \\
    \hline
    QP & 42 & 45 & 48 & 50 & 52 & 55 & 58 & 60 & 62  \\
    \hline
    bpf & 255 & 203 & 169 & 153 & 141 & 130 & 123 & 121 & 112  \\
    $10^{-3}$bpp & 3.89 & 3.1 & 2.58 & 2.33 & 2.15 & 1.98 & 1.88 & 1.85 & 1.71  \\
    KBps & 0.93	&0.74&	0.62&	0.56&	0.52&	0.48&	0.45&	0.44&	0.41  \\
    PSNR(Y) & 32.71 & 31.01 & 29.26 & 28.12 & 27.00 & 25.47 & 24.11 & 23.32 & 22.55  \\
    \hline
    \end{tabular}
\end{center}
\end{table*}

\begin{table*}[htp]
\small
\renewcommand\arraystretch{1.15}
\vspace{-0.4cm}
\begin{center}
\caption{Experimental results of our method under different QP values and sampling intervals. Mode: LDP.}
\label{tab:ours_ldp}
\setlength\tabcolsep{8pt}
\vspace{0.2cm}
    \begin{tabular}{|c|c|c|c|c|c|c|c|c|c|c|c|}
    \hline
    QP & \multicolumn{2}{c|}{sampling-interval} & 5 & 10 & 20 & 30 & 40 & 50 & 60 & 80 & 100  \\
    \hline 
    & \multicolumn{2}{c|}{non-key-frame / Bpf} & 5.358 & 5.079 & 4.964 & 4.922 & 4.907 & 4.895 & 4.887 & 4.876 & 4.873  \\
    \hline
    
    \multirow{5}{*}{27} & \multicolumn{2}{c|}{key-frame / bpf} & 2848&	3798&	4941&	5842&	6225&	6822&	7568&	8579&	9497 \\
    \cline{2-12}
    & \multirow{3}{*}{average-bitrate} & bpf  & 603.9&	416.4&	284.8&	232.8&	193.9&	174.8&	164.6&	145.8&	133.6 \\
    & & $10^{-3}$bpp & 9.21	&6.35	&4.35	&3.55&	2.96&	2.67&	2.51&	2.22&	2.04 \\
    & & KBps  & 2.21& 1.52&	1.04&	0.85&	0.71&	0.64 &	0.6&	0.53&	0.49\\
    \cline{2-12}
    &  \multicolumn{2}{c|}{PSNR(Y)} & 37.85&	36.23&	34.51&	33.40&	32.71&	32.13&	31.54&	31.11&	30.62 \\
    \hline
    
    \multirow{5}{*}{30} & \multicolumn{2}{c|}{key-frame / bpf} & 1910&	2575&	3387&	4062&	4364&	4822&	5377&	6190&	6889 \\
    \cline{2-12}
    & \multirow{3}{*}{average-bitrate} & bpf & 416.3&	294.1&	207.1&	173.5&	147.4&	134.8&	128.1&	115.9&	107.5 \\
    & & $10^{-3}$bpp & 6.35	&4.49&	3.16&	2.65&	2.25&	2.06&	1.95&	1.77&	1.64 \\
    & & KBps  & 1.52&	1.08&	0.76 &	0.64&	0.54&	0.49&	0.47 &	0.42&	0.39 \\
    \cline{2-12}
    &  \multicolumn{2}{c|}{PSNR(Y)}& 36.85&	35.52&	33.97&	33.02&	32.32	&31.83&	31.30&	30.92	&30.44 \\
    \hline
    
    \multirow{5}{*}{32} &\multicolumn{2}{c|}{key-frame / bpf} & 1452&	1969&	2626&	3171&	3430&	3800&	4272&	4941&	5516 \\
    \cline{2-12}
    & \multirow{3}{*}{average-bitrate} & bpf  &324.7&	233.5&	169	&143.8&	124	&114.4&	109.6&	100.3	&93.8 \\
    & & $10^{-3}$bpp & 4.95&	3.56&	2.58&	2.19&	1.89&	1.75&	1.67&	1.53&	1.43 \\
    & & KBps  &1.19    &   	0.86 &	0.62&	0.53&	0.45&	0.42&	0.4	&0.37&	0.34 \\
    \cline{2-12}
    &  \multicolumn{2}{c|}{PSNR(Y)}&36.10&	34.89&	33.53&	32.58	&32.09	&31.60&	31.05&	30.62&	30.30 \\
    \hline
    
    \multirow{5}{*}{34} & \multicolumn{2}{c|}{key-frame / bpf} & 1155&	1575&	2114&	2563&	2784&	3100&	3526&	4057&	4559 \\
    \cline{2-12}
    & \multirow{3}{*}{average-bitrate} & bpf  & 265.3&	194.1&	143.4	&123.5&	107.9&	100.4	&97.2&	89.2&	84.2 \\
    & & $10^{-3}$bpp & 4.05	&2.96&	2.19&	1.88&	1.65&	1.53&	1.48&	1.36&	1.28 \\
    & & KBps  & 0.97    &	0.71	&0.53&	0.45&	0.4&	0.37 &	0.36& 	0.33&	0.31 \\
    \cline{2-12}
    &  \multicolumn{2}{c|}{PSNR(Y)} & 35.30&	34.31&	33.04	&32.19&	31.70&	31.28&	30.80&	30.53&	30.10 \\
    \hline

    \multirow{5}{*}{36} & \multicolumn{2}{c|}{key-frame / bpf} & 925	&1261	&1700	&2088&	2264&	2524&	2881&	3325&	3760 \\
    \cline{2-12}
    & \multirow{3}{*}{average-bitrate} & bpf & 219.3&	162.7&	122.7&	107.7&	94.9&	88.9&	86.5&	80.1&	76.2\\
    & & $10^{-3}$bpp & 3.35&2.48&	1.87&	1.64&	1.45&	1.36&	1.32&	1.22&	1.16 \\
    & & KBps  & 0.8	&0.6&	0.45&	0.39&	0.35&	0.33&	0.32 &	0.29&	0.28 \\
    \cline{2-12}
    &  \multicolumn{2}{c|}{PSNR(Y)}& 34.45&	33.60&	32.46&	31.72&	31.28&	30.92&	30.48&	30.14&	29.88 \\
    \hline
    
    \multirow{5}{*}{38} & \multicolumn{2}{c|}{key-frame / bpf} & 744	&1016&	1379&	1706&	1862&	2072&	2377&	2756&	3117 \\
    \cline{2-12}
    & \multirow{3}{*}{average-bitrate} & bpf &183.1&	138.2&	106.7	&94.9&	84.8&	79.8&	78.1&	73&	69.8 \\
    & & $10^{-3}$bpp &2.79&	2.11&	1.63&	1.45&	1.29&	1.22&	1.19&	1.11&	1.07 \\
    & & KBps  & 0.67 &0.51	&0.39&	0.35&	0.31&	0.29&	0.29 &	0.27&	0.26 \\
    \cline{2-12}
    &  \multicolumn{2}{c|}{PSNR(Y)} &33.57	&32.82&	31.83&	31.09&	30.79	&30.40&	30.06&	29.85&	29.54 \\
    \hline
   
    \end{tabular}
\end{center}
\end{table*}

\end{document}